\documentclass[letterpaper, 10 pt, conference]{ieeeconf}  

\IEEEoverridecommandlockouts                              

\usepackage{comment}
\usepackage{graphics} 
\usepackage{epsfig} 
\usepackage{times} 
\usepackage{amsmath} 
\usepackage{amssymb}  
\usepackage{bbold}  
\usepackage[space]{cite}
\usepackage{xcolor}
\usepackage{enumerate}

\newtheorem{remark}{Remark}
\newtheorem{problem}{Problem}
\newtheorem{theorem}{Theorem}
\newtheorem{assumption}{Assumption}
\newtheorem{corollary}{Corollary}
\usepackage{version}
\includeversion{arxiv}
\excludeversion{cdc}

\title{\LARGE \bf Reactive power flow optimization in AC drive systems}

\author{Sanjay Chandrasekaran$^{1,3}$, Catalin Arghir$^{1}$,  Pieder Jörg$^{2}$,  Florian Dörfler$^{3}$ and  Silvia Mastellone$^{1}$
\thanks{This work was supported by the Swiss National Science Foundation under NCCR Automation, grant agreement 51NF40\_180545}
\thanks{$^{1}$Sanjay Chandrasekaran, Catalin Arghir and Silvia Mastellone are with the Institute of Electric Power Systems, University of Applied Sciences Northwest Switzerland, Windisch, Switzerland. \texttt{ sanjay.chandrasekaran@fhnw.ch, catalin.arghir@fhnw.ch, silvia.mastellone@fhnw.ch}}
\thanks{$^{2}$Pieder Jörg is with ABB Motion, ABB Switzerland Ltd., Turgi, Switzerland.
\texttt{ pieder.joerg@ch.abb.com}}
\thanks{$^{3}$Sanjay Chandrasekaran and Florian Dorfler are with the Automatic Control Lab (IfA) at ETH Zurich, Switzerland. 
\texttt{schandraseka@ethz.ch, dorfler@ethz.ch}}}

\begin{document}

\maketitle
\thispagestyle{empty}
\pagestyle{empty}

\begin{abstract}

This paper explores a limit avoidance approach in the case of modulation (input) and current (output) constraints with the aim of enhancing system availability of AC drives. 
Drawing from the observation that, in a certain range of reactive power, there exists a trade-off between current and modulation magnitude, we exploit this freedom and define a constrained optimization problem. 
We propose two approaches: one in the form of an activation-function, and another that uses online feedback optimization to set the reactive power dynamically. Both methods compromise reactive power tracking accuracy for increased system robustness.
Through a high-fidelity simulation, we compare the benefits of the two methods, highlighting their effectiveness in industrial applications.
\end{abstract}

\section{INTRODUCTION}

Medium voltage (MV) drives are widely used in industrial applications, including heat pumps, compressors or wind turbines, where they enable efficient and precise control of large electrical machines at variable speed \cite{rodriguez2007multilevel, zargari2017guide}.  
These drives often employ voltage-source converters (VSCs) to regulate power flow and maintain stability and flexibility of operation. However, during some transient events such as overload or grid disturbances, the drive system can experience overcurrent and overmodulation, leading to degraded performance, and in extreme cases, system shutdown.
In applications such as oil and gas, the unplanned interruption of the process results in high financial loss. Thus, system availability requirements are sometimes prioritized over performance and efficiency. 

Reactive power in drive systems is 
typically regulated to zero \cite{teodorescugrid} in order to minimize losses. 
However, in megawatt-scale applications, MV drives with active front-ends can significantly influence grid behavior, making reactive power management a crucial factor \cite{shakerighadi2023overview}. The design of AC drive systems generally involves a trade-off between efficiency and robustness to grid conditions, allowing some margin for reactive power needed for Static Synchronous Compensator (STATCOM) features \cite{Siemens2019}. In applications such as wind energy, VSCs actively regulate the point of common coupling (PCC) voltage or may participate in grid-forming control \cite{gomis2022grid}. In other cases, they provide ancillary services by allowing the reactive power set point to be determined by the grid operator or by implementing a droop between reactive power and voltage. 

However, the VSC has a limited range of supplying reactive power. Its constraints arise in the static case, due to design limitations, and the dynamic case, as a function of instantaneous active power, DC-link voltage, and grid conditions \cite{schneeberger2023sos, 5446354}. 
We propose a constraint-aware dynamic adaptation of the reactive power set-point that accounts for external demands while respecting system constraints.

In this study, we investigate two competing methods, an activation-function approach and an optimization-based technique leveraging Online Feedback Optimization (OFO). Both methods operate as outer-loops to existing controllers with anti-windup limiters, preventing them from entering the saturated mode of operation whenever possible. By integrating these strategies, we enhance both transient and steady-state performance when compared to state-of-the-art current limiters \cite{10603443, desai2024saturation, 9807120}.

In the context of power systems, various types of strategies for limiting converter current are employed \cite{abolmasoumi2024state,chen2023impact, bernal2024complex, murad2021modeling}. We aim to leverage the relationship between anti-windup and the projected gradient descent mechanism found in OFO \cite{picallo2022adaptive}. In a similar scenario \cite{zagorowska2024tuning}, OFO is used to steer fast nonlinear dynamics. In \cite{ortmann2023deployment}, it is used to optimize reactive power flow in a distribution grid by prescribing the inverter set-points while enforcing voltage constraints. In our work, we take a deeper look into the single-inverter case and enforce constraints with the aim of internal stability.

    The rest of the paper is organized as follows: Section \ref{prbState} formulates the problem, describing the system model, control assumptions, and constraints. Section \ref{ctrlPrb} goes over the control design, while Section \ref{convGar} presents the OFO convergence guarantees. Section \ref{expRes} provides simulation results that illustrate performance improvements achieved by the proposed methods. Finally, Section \ref{Concl} concludes the paper with a discussion of key findings.

\section{Problem statement}
\label{prbState}
We consider a power conversion system consisting of a grid-side converter, a DC-link capacitor, a motor-side converter, an electrical machine, and a driveline shaft attached to a motoring or generating load. 
%
The main role of the grid-side converter, is to deliver active power by regulating the DC-link voltage, and thus allow the motor-side converter to supply the load under a wide range of external conditions.

Our main goal is to design an outer loop for the reactive-power component of the grid current reference 
to avoid overmodulation or overcurrent limits and preserve DC-link regulation.
%
%
Our proposal is to exploit the trade-off between converter current and modulation constraints and override the external reactive power set-point. \textcolor{black}{The problem setup is developed under the following assumptions}.



\begin{assumption} \label{Modellingassumptions}
The drive system is modeled based on the following set of assumptions:
    \begin{enumerate}[i.]
    \item Changes in set-points, load torque, and the grid voltage have limited bandwidth;
    \item The inner loop of the motor-side converter is assumed to have sufficiently fast bandwidth such that, together with the motor dynamics, it can be abstracted out of the overall model by assuming that the motor torque is perfectly tracking its set-point.
\end{enumerate}
\end{assumption}
\textcolor{black}{Assumption \ref{Modellingassumptions}(i) provides a limit on the variation of external disturbances. Assumption \ref{Modellingassumptions}(ii) implies that energy is preserved from the drive shaft to the DC-link. The second assumption is motivated by the fact that most modern MV motor controllers adopt a form of direct Model predictive control (MPC), or Direct torque control (DTC) \cite{geyer2008model, tiitinen1996next}, to track the torque set point with high bandwidth. 
}

%
%
%


\subsection{Drive system model}

In the rest of this section, we explain a typical industrial drive and its core control system. We propose to use the following energy-preserving, average-switch model of a back-to-back drive system:
\begin{subequations}
\label{system_eq}
\begin{align}
    M\dot{{w}} &= -D{w} + {\tau}_m{} - \tau_l \label{omegadynamics}
    \\
    C_{dc}\dot{v}_{dc} &= -G_{dc}v_{dc} - \tfrac{{{w}}}{v_{dc}}{\tau}_m{} + m_g{}^\top i_g
    \label{vdcdynamics}\\
    L_g\dot{i}_g &= - Z_g i_g  + v_g - m_g{}v_{dc} \,, \label{igdynamics}
\end{align}
\end{subequations}
where $M,D$ are the moment of inertia and viscous damping of the driveshaft, $w$ its angular velocity, ${\tau}_m$ is the motor air-gap torque (control input), $\tau_l$ is the driveshaft load torque (unmeasured disturbance), $v_{dc}$ is the DC-link voltage, $C_{dc}$ its capacitance and $G_{dc}$ its parallel conductance. Furthermore, $m_g\in\mathbb{R}^2$ is the grid-side converter $dq$-modulation vector (control input), $L_g,R_g>0$ phase inductance and resistance, respectively, $i_g\in\mathbb{R}^2$ is the grid current (towards the converter) and $v_g\in\mathbb{R}^2$ is the voltage in $dq$-coordinates (measured disturbance). Finally, the grid impedance (in $dq$-coordinates) is given by $ Z_g = \begin{bmatrix}R_g & -\omega_0L_g \\ \omega_0L_g & R_g\end{bmatrix}$.



{We use the well established notions of instantaneous power \cite{o2019geometric} and the power-invariant $dq$-frame transformation as} $x_{abc} \mapsto x_{dq0} = T_{\theta_g} T_{\alpha\beta\gamma} x_{abc}$
such that $T_{\alpha\beta\gamma}^{-1} = T_{\alpha\beta\gamma}^\top$ and $T_{\theta_g}^{-1} = T_{\theta_g}^\top$, and assume that the grid voltage angle $\theta_g$ is measured. Moreover, by virtue of the third-harmonic injection, the maximum modulation amplitude in $dq$-coordinates becomes $\tfrac{1}{\sqrt{2}}$.

\subsection{Drive system objectives}
\label{control_section} 
 
We consider the following tracking objectives, 
\begin{align}
    {w} &\rightarrow {w}^\mathrm{ref}  
    \label{trackingobjective_w}\\
    v_{dc} &\rightarrow v_{dc}^\mathrm{ref} 
    \label{trackingobjective_vdc}\\
    Q_g = v_{g}^\top J i_g &\rightarrow Q_g^\star \,\text{ , where }\, J = \begin{bmatrix} 0 & -1 \\ 1 & 0 \end{bmatrix} \, \label{trackingobjective_Qg} \,,
\end{align}
while rejecting external disturbance $\tau_l, v_g$. Moreover, $|{w}^\mathrm{ref}|\leq w_\mathrm{max}$ and $|Q_g^\star|\leq Q_\mathrm{max}$ are slowly changing external set-points. The system is subject to fast changing loads $|\tau_l|\leq \tau_\mathrm{max}$ and occasional dips in $\|v_g\|\leq v_{g,\mathrm{max}} = \alpha_v v_{g,\mathrm{nom}}$ with e.g. $\alpha_v = 1.1$ representing a grid code overvoltage limit. We also consider some margin for producing reactive power 
\begin{align}\label{maxpower}
    \tau_\mathrm{max}w_\mathrm{max} = P_{g,\mathrm{max}} = \tfrac{1}{\alpha_q} \sqrt{P_{g,\mathrm{max}}^2 + Q_{g,\mathrm{max}}^2} \,,
\end{align}
where e.g. $\alpha_q = 1.2$ is a de-rating factor, such that the maximum current amplitude becomes $i_{g,\mathrm{max}} = \tfrac{\alpha_q P_{g,\mathrm{max}}}{v_{g,\mathrm{nom}}}$. 

\subsection{Inner loop definition}

The standard practice involves a cascade of PI controllers with anti-windup. We consider the speed control as
\begin{subequations}
\label{speedcontrol_eq}
\begin{align}\label{speed_PI}
    \dot{x}_m &=  {w}-{w}^\mathrm{ref} 
    \\
    \tau_m &= \underset{|\cdot| \leq \tau_\mathrm{max}}{\mathrm{sat}}[ -K_{p,m}({w}-{w}^\mathrm{ref} ) - K_{i,m}x_m]\,,
\end{align}
\end{subequations}
where $K_{p,m} = 2 \zeta_{m} \omega_{m} M$, and $K_{i,m} = \omega_{m}^{2} M$, with tunable $\omega_{m}$ and $\zeta_{m}$, and $\underset{|\cdot|\leq a}{\mathrm{sat}}[\cdot]$ represents 
bounding the argument in absolute value (or in norm for the multivariable case).  

We define the DC-link control as

\footnotesize
\begin{subequations}
\label{dccontrol_eq}
\begin{align}\label{vdc_PI}
    \dot{x}_{dc} &= v_{dc}-v_{dc}^\mathrm{ref}
    \\
    P_g^\star &= \underset{|\cdot| \leq P_{g,\mathrm{max}}}{\mathrm{sat}}[ (-K_{p,dc}(v_{dc}\!-\!v_{dc}^\mathrm{ref}) \!-\! K_{i,dc}x_{dc})v_{dc}^\mathrm{ref} + \tau_m {w}^{\rm ref}],
\end{align}         
\end{subequations}
\normalsize where $K_{p,dc} = 2 \zeta_{dc} \omega_{dc} C_{dc}$, and $K_{i,dc} = \omega_{dc}^{2} C_{dc}$, with tunable $\omega_{dc}$ and $\zeta_{dc}$. Moreover, the reference for the motor torque is used as feed-forward from the speed controller. 


For the grid-side current reference, we consider a circular current limiter, as described below
\begin{align}
   i_{g}^\star = \underset{\|\cdot\| \leq i_{g,\mathrm{max}}}{\mathrm{sat}}\tfrac{1}{\|v_g\|^2}\begin{bmatrix}v_g^\top \\ v_g^\top J\end{bmatrix}\begin{bmatrix}P_g^\star \\ Q_g^\star\end{bmatrix} \label{igstar_ssmap} \,.
\end{align}

The PI controller for current is then given by:
\begin{subequations}
\label{currentcontrol_eq}
\begin{align}\label{ig_PI}
\dot{x}_{g} &= i_{g}-i_{g}^\star
    \\
m_g &= \underset{\|\cdot\| \leq \tfrac{1}{\sqrt{2}}}{\mathrm{sat}}[\tfrac{1}{v_{dc}^\mathrm{ref}}(v_g - Z_gi_g^\star + K_{p,g}(i_{g}-i_{g}^\star) + K_{i,g}{{x_{g}}})] \label{mg_equation} \,,
\end{align} 
\end{subequations}
where $K_{p,g} = 2 \zeta_{g} \omega_{g} L_g$, and $K_{i,g} = \omega_{g}^{2} L_g$, with tunable $\omega_{g}$ and $\zeta_{g}$. In this scenario we take $\zeta_m, \zeta_{dc}, \zeta_g = 1 $, and set both $\omega_{dc}, \omega_m$ much smaller than current control bandwidth $\omega_g$, enforcing time-scale separation. 





\section{Control problem}
\label{ctrlPrb}

We are now ready to state our main control problem.
\begin{problem}\label{controlproblem}
    Consider the system \eqref{system_eq} in feedback with \eqref{speedcontrol_eq}-\eqref{currentcontrol_eq}. The objective is to prescribe a feedback law on $Q_g^\star$ such that the limits corresponding to the constraints $\|i_g\|\leq i_{g,\mathrm{max}}$ and $\|m_g\| \leq \tfrac{1}{\sqrt{2}}$ are avoided at trade-off with an external set-point for reactive power. 
\end{problem}

Our intuition is based on the fact that when ${Q}_g^\star$ is set to zero, the current magnitude (and therefore power losses) are minimized, and when ${Q}_g^\star$ is set to ${Q}_{g}^\textit{mm} = \tfrac{\omega_0L_g \|v_g\|^2}{\|Z_g\|^2}$,
the modulation amplitude is minimized, and therefore robustness is improved. The latter is seen by inserting \eqref{igstar_ssmap} into \eqref{mg_equation} and taking the norm

\begin{footnotesize}
\begin{align*}
\|m_g\|^2{v_{dc}^\mathrm{ref}}^2 &= \|v_g\|^2 - 2R_g P_g^\star - 2\omega_g L_g Q_g^\star + \tfrac{\|Z_g\|^2}{\|v_g\|^2}(P_g^{\star2} + Q_g^{\star2})\,,
\end{align*}
\end{footnotesize}
and noticing that the gradient is zero at ${Q}_{g}^\textit{mm}$,
\begin{align}
\nabla_{Q_g^\star} {\|m_g\|^2} &= \tfrac{2}{{v_{dc}^\mathrm{ref}}^2 }(-\omega_g L_g + \tfrac{\|Z_g\|^2}{\|v_g\|^2}Q_g^{\star})\,.
\end{align}
\begin{figure}[!ht]
\centering{
\includegraphics[scale=0.26]{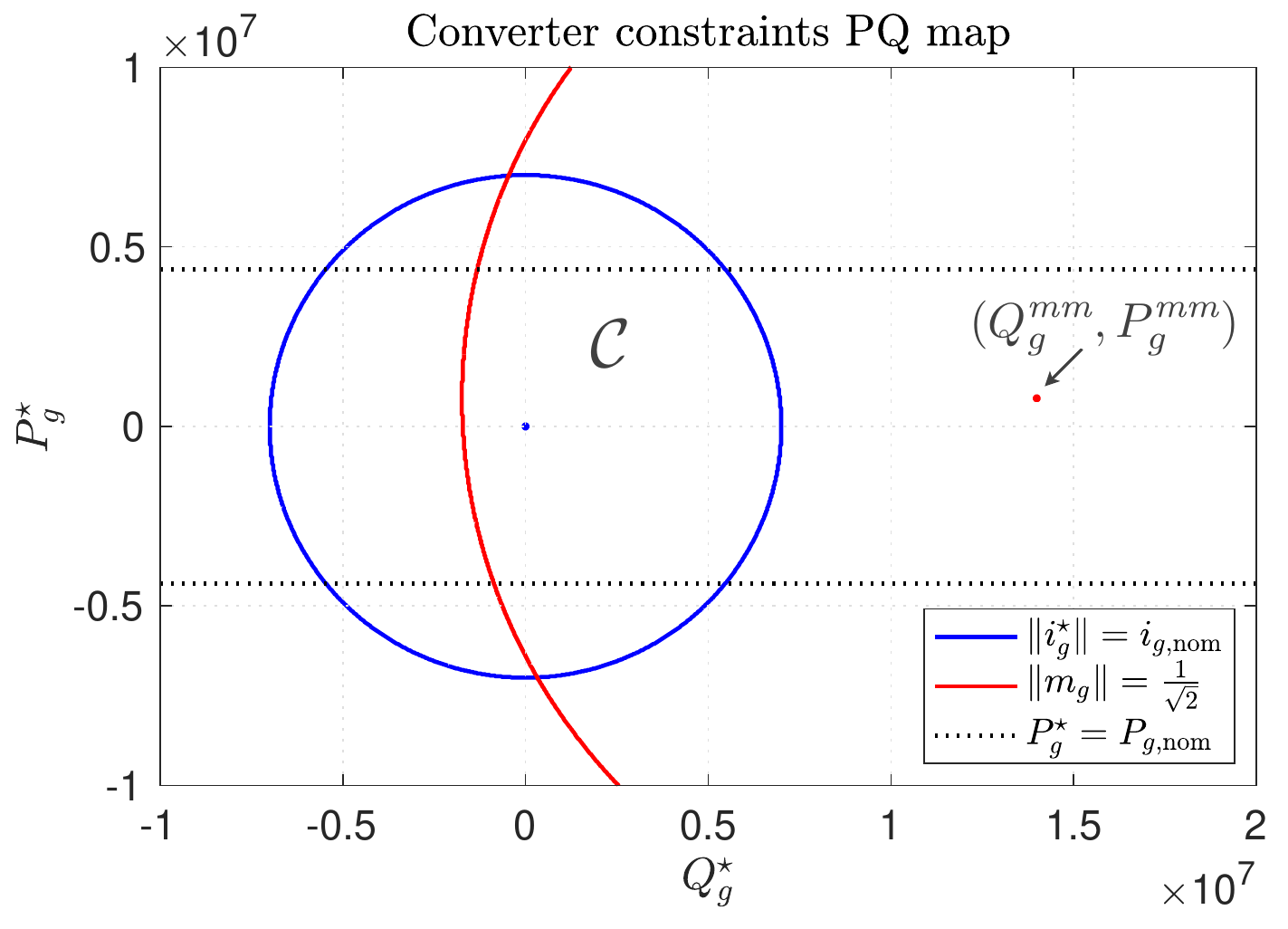}
\caption{The PQ diagram for the considered drive system at nominal conditions, showing the admissible $Q_g$ set-point as constrained by the active power demand.}	\label{Fig: pqmap}}
\end{figure}


As seen in Fig. \ref{Fig: pqmap}, for our converter design, the trade-off between current and modulation magnitude is present in most of the admissible range of operation. Note that it is easier for the VSC to inject positive rather than negative (or capacitive) reactive power into the grid, and, in order for the drive system to act as a STATCOM, a sufficient DC-bus margin must be provided. 

\subsection{Activation-function approach}

Our first proposal is to use a low-pass filter on the external reference which is deterred to safety values when the corresponding limits are reached
%
\begin{align}
\label{AF_eqn}
    \dot{Q}_g^\star =&~ -\omega_{q}({Q}_g^\star - {Q}_g^\mathrm{ref}) \notag
    \\
    &~- \kappa_1\Gamma_1(i_g^\star){Q}_g^\star - \kappa_2\Gamma_2(m_g)({Q}_g^\star - {Q}_{g}^\textit{mm})\,.
\end{align}
\normalsize
Here $\omega_{q} > 0$ is the speed at which the controller tracks the unconstrained set-point ${Q}_g^\mathrm{ref}$. Furthermore, $\kappa_1, \kappa_2 > 0$ are gains on the sensitivity to constraint violations for $i_g^\star$, respectively $m_g$, and the two positive functions
\begin{align}
    \Gamma_1(i_g^\star) &= \|i_g^\star\| - \textit{min}(\|i_g^\star\|, i_{g,\mathrm{max}})
    \\
    \Gamma_2(m_g) &= \|m_g\| - \textit{min}(\|m_g\|, \tfrac{1}{\sqrt{2}})\,,
\end{align}
are activations of the gradients that drive $Q_g^\star$ to the values that mitigate the corresponding constraint violations. 

    \begin{figure}[!ht]
\centering{
\includegraphics[scale=0.24]{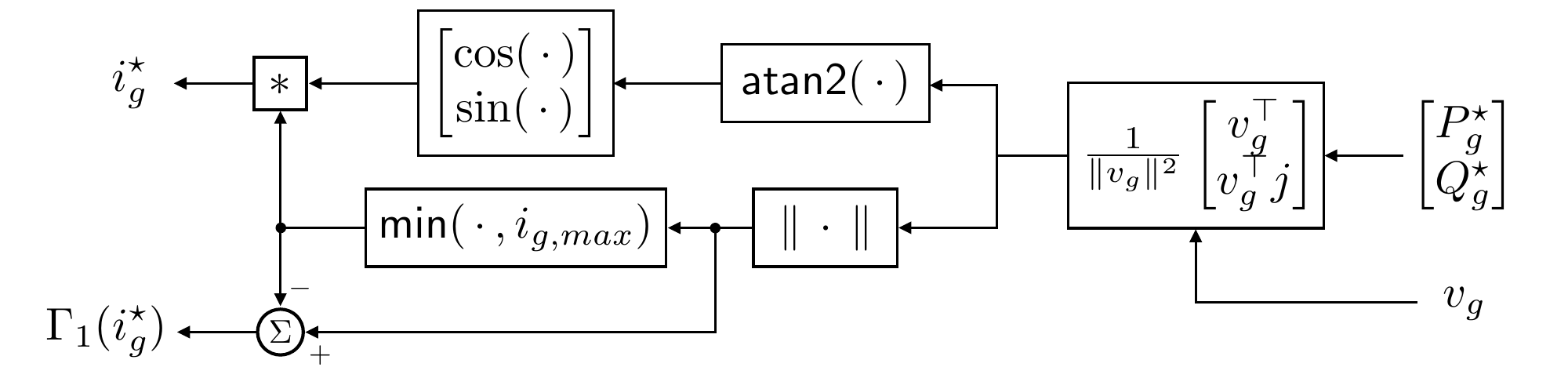}
\caption{Block diagram of the circular current reference limiter with $\Gamma_1$.}	\label{Fig: igsbl}}
\end{figure}

\begin{figure}[!ht]
\centering{
\includegraphics[scale=0.24]{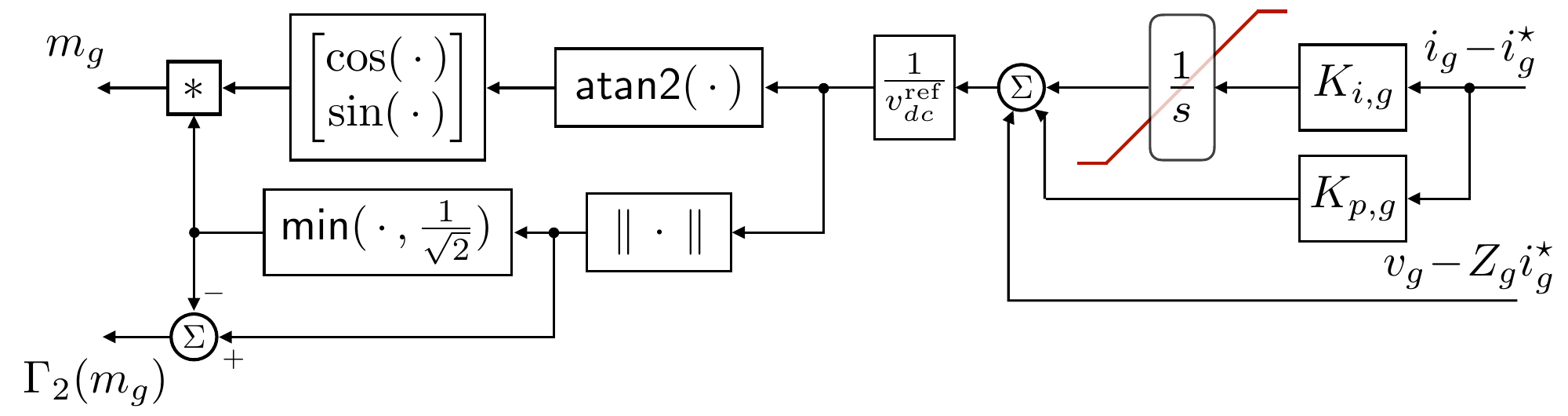}
\caption{Block diagram of the circular modulation limiter with $\Gamma_2$.}	\label{Fig: mgbl}}
\end{figure}

Note that, in Fig. \ref{Fig: mgbl}, the integrator saturation block implements, for simplicity, a rectangular limiter on each $dq$-component circumscribing the circular limit $\|m_g\| \leq \tfrac{1}{\sqrt{2}}$. The role of $\Gamma_2$ is to keep the modulation magnitude bounded by indirectly driving the integrator away from saturation.

\begin{remark}\label{prioritizingconstraints}
    It is important to assign higher priority to the modulation constraint by tuning $\kappa_1 \ll \kappa_2$, as the converter loses its ability to track $i_g^\star$ and therefore is unable to limit the current when the modulation constraint limit is reached.
\end{remark}

\begin{remark}\label{balanced-limit}
An alternative to the circular limiter may be one with active-power priority \cite{9944962}. However, our approach has the advantage of being sensitive, in addition, to the (asymmetric) saturation of the modulation magnitude. 
\end{remark}

Despite its simplicity and compatibility with existing saturation mechanisms, this method lacks formal convergence guarantees, which leads to our next proposal.

\subsection{OFO: Problem Formulation}

An alternative view to preserving the drive system operation is to iteratively compute a suitable set point $Q_g^\star$ satisfying current and modulation constraints. Thus, a steady-state optimization algorithm such as OFO \cite[Section 3]{Adrian} is convenient. 
We define the optimization problem as
\begin{subequations}
\label{optimizationproblem}
\begin{align} 
    \underset{Q_g^\star}{\mathrm{minimize}} &\quad   \varphi(Q_g^\star,i_g^\star) \label{opt_cost} \\
    \mathrm{s.t.}  & \quad i_g^\star = h(Q_g^\star,d) \label{ssmap} \\
    & \quad Q_g^\star \in \mathcal{C} \label{constraintspace},
\end{align}
\end{subequations}
where, $\varphi(i_g,Q_g^\star)= \frac{1}{2}(\lVert i_g \rVert^2 + \gamma\lVert Q_g^\star-Q_g^{\rm ref} \rVert^2) $, with $\gamma$ being a non-negative trade-off parameter between minimizing the current $i_g$ and the deviation of the reactive power set-point $Q_g^\star$ from its external reference $Q_g^{\rm ref}$. Since the natural choice for current minimization is minimizing $Q_g^\star$, we construct a composite cost function including the term $\lVert Q_g^\star-Q_g^{\rm ref} \rVert^2$ as a soft constraint that restricts $Q_g^\star$ from deviating significantly from an external reference $Q_g^{\rm ref}$. The steady-state map $i_g^\star = h(Q_g^\star,d(k))$ is described in \eqref{igstar_ssmap} 
where $d(k) = (P_g^\star(k), v_g(k))$
represents a time-varying disturbance vector. The constraints $\mathcal{C}(k)$ represent limits on the current and modulation amplitude at timestep $k$. Using \eqref{igstar_ssmap} and \eqref{mg_equation}, the current constraint $\lVert i_g^\star \rVert^2 \leq i_{\rm g,\mathrm{max}}^2$ and modulation constraint $\lVert m_g \rVert^2 \leq \tfrac{1}{2}$ at each operating point are written in terms of $Q_g^\star$ as follows, respectively:

\footnotesize
\begin{align}
    \mathcal{C}(k) = \Bigg\{ Q_g^\star:\, & P_g^{\star2}(k) + Q_g^{\star2} \leq \lVert v_g(k) \rVert^2i_{g, \rm max}^2, \nonumber \\
    & \big\lVert v_g(k) - Z_gh(Q_g^\star,d(k))\rVert^2 \leq \frac{1}{2}(v_{dc}^\mathrm{ref})^2 \Bigg\} \label{Qconstraints}
\end{align}
\normalsize
\begin{remark}\label{convexityremark}
Since the cost function $\varphi(Q_g^\star,i_g)$ is strongly convex in $Q_g^\star$, and steady-state map $h(Q_g^\star,d(k))$ affine with respect to $Q_g^\star$ and $\mathcal{C}(k)$ convex at every time instant $k$, the resulting optimization problem in \eqref{optimizationproblem} is strongly convex and admits a unique solution trajectory $\{i_g^\star(k),Q_g^\star(k)\}_{k=0}^N$ for every unique load disturbance sequence $\{d(k)\}_{k=0}^N$.
\end{remark}

\subsection{OFO: Problem Solution}
Using OFO, we approach Problem \ref{controlproblem} with a feedback controller based on the following projected gradient descent algorithm \cite[Section II.D]{picallo2022adaptive}:
\begin{equation}\label{PGD}
    Q_g^\star(k+1) = \Pi_{\mathcal{C}(k)}[Q_g^\star(k) - \mu(k)\zeta(k)\Phi(Q_g^\star(k),i_g(k+1))],
\end{equation}
where $\mu$ is the learning rate, $\Pi_{\mathcal{C}(k)}[\cdot]$ is the Euclidean projection onto set $\mathcal{C}(k)$, and the composite gradient
\begin{equation*}
    \Phi(Q_g^\star,i_g) = \nabla_{Q_g^\star}\varphi(Q_g^\star,i_g) + \nabla_{Q_g^\star}h(Q_g^\star,d(k))^\top\nabla_{i_g}\varphi(Q_g^\star,i_g)
\end{equation*}
results from the chain-rule of differentiating the cost $\varphi(Q_g^\star,h(Q_g^\star,d(k))$. As in every OFO algorithm, we replace the steady-state value of $i_g$ in \eqref{igstar_ssmap} with its real-time measurement $i_g(k+1)$ assuming that $i_g$ has reached a local steady-state \cite{ortmann2023deployment}. In practice, this time-scale separation between the system and the controller can be enforced by introducing a periodic trigger \cite[Section III.B]{chandrasekaran2024} that produces a new OFO solution every $m$ time steps (i.e. $\zeta(k) = 1$ if $k=l m$, where $l=\{0,1,2\ldots\}$ and $\zeta(k)=0$ if $k\neq l m$). Thus, the sampling time of the OFO and the system are related as $T_s=mT_c$, with $T_s,T_c$ being the sampling time of the OFO and plant, respectively.

\section{Convergence Guarantees}
\label{convGar}

Here, we provide convergence guarantees for the closed-loop system, with the inner-loop \eqref{omegadynamics}-\eqref{igdynamics} regulated using the PI controllers \eqref{speed_PI},\eqref{vdc_PI},\eqref{ig_PI} and the outer-loop controlled by the OFO based on the projected gradient descent algorithm in \eqref{PGD}. We make the following assumption regarding the inner-loop and the disturbances, respectively:
\begin{assumption}\label{PI_assumption}
    The PI controllers \eqref{speed_PI},\eqref{vdc_PI},\eqref{ig_PI} are tuned such that the inner-loop is locally exponentially stable for a constant input $Q_g^\star$ and disturbance $d$, i.e. $\exists \ C_1, C_2:\lVert i_g(k) - h(Q_g^\star,d)\rVert \leq C_1\lVert i_g(0) - h(Q_g^\star,d)\rVert e^{-C_2k}$
\end{assumption}
\begin{assumption}\label{disturbance_assumption}
    The disturbance $d(k)$ is piece-wise constant with respect to the sampling rate of OFO, i.e. $d(l m)=d(l m + 1) = \ldots = d(l m + m-1), \forall l \in \{0,1,2,\ldots\}$.
\end{assumption}
Assumption \ref{disturbance_assumption} quantitatively describes bandwidth limitations on the disturbances and load (see Assumption \ref{Modellingassumptions}(ii)). \textcolor{black}{With Assumption \ref{PI_assumption} and using \eqref{system_eq}, \eqref{igstar_ssmap}, we obtain $P_g^\star = D(\omega^{\rm ref})^2+G_{dc}(v_{dc}^{\rm ref})^2 + \omega^{\rm ref}\tau_l$ at steady-state. Since the variations in the external reference for angular frequency $\omega^{\rm ref}$, DC-link voltage $v_{dc}^{\rm ref}$, load torque $\tau_l$ and grid-side disturbance $v_g$ occur at slower time-scales, it is justified to assume that these disturbances remain constant between two OFO sampling times (which qualitatively operate at a much faster time-scale compared to the disturbances).}

To analyze convergence guarantees, we use the metric $\psi(k) =|Q_g^\star(k) - \bar{Q}_g^\star(k)|$ that describes the distance of the OFO solution from its global optimum $\bar{Q}_g^\star(k)$ at time $k$. 

\begin{theorem}\label{maintheorem}
    Let Assumptions \ref{Modellingassumptions},\ref{PI_assumption}, \ref{disturbance_assumption} hold. For any learning rate $\mu(k) \in \Big(0,\frac{2\lVert v_g(k)\rVert^2}{1 + \gamma\lVert v_g(k)\rVert^2}\Big)$, the sequence of $Q_g^\star(k)$ generated using \eqref{PGD} satisfies the following recursive relation:
    
    \footnotesize
    \begin{align}
        \psi((l+1)m) \leq & \epsilon(l m) \psi(l m) + \underset{\mathcal{T}_1}{\underbrace{|\Delta \bar{Q}_g^\star(l m)|}} + \nonumber\\
        & \underset{\mathcal{T}_2}{\underbrace{C_1C_3\lVert v_g(l m)\rVert\lVert i_g(l m) - i_g^\star(l m) \rVert e^{-C_2}}}, \label{convergence}
    \end{align}
    \normalsize where, $\Delta \bar{Q}_g^\star(k)= \bar{Q}_g^\star(k)-\bar{Q}_g^\star(k+1)$, $\epsilon(k)= 1-\mu(k)(\gamma + \lVert v_g(k) \rVert^{-2})$, $i_g^\star(k)= h(Q_g^\star(k),d(k))$, the constants $C_1,C_2$ resulting from Assumption \ref{PI_assumption} and $C_3$ from using $\mu(k) = C_3\lVert v_g(k)\rVert^2$, $C_3\in(0,2/(1+\gamma v_{g,\mathrm{max}}^2))$. 
\end{theorem}
\begin{arxiv}
    \begin{proof}
        Refer to Appendix \ref{maintheoremproof}
    \end{proof}
\end{arxiv}
\begin{proof}
        \textcolor{black}{Refer to the proof in the extended version of our paper in} \cite{chandrasekaran2025reactive}.
\end{proof}

The above theorem shows that with an appropriate choice of $\mu(\cdot)$ such that $|\epsilon(\cdot)|<1$, the distance of the OFO trajectory from its optimal solution is contracting, with the offset terms $\mathcal{T}_1$ accounting for the difference in the optimal reactive power between two trigger time instances and $\mathcal{T}_2$ accounting for the disturbance $v_g(\cdot)$ and the distance of $i_g(\cdot)$ from the new steady-state point $i_g^\star(\cdot)$. 

\begin{corollary}\label{asymptoticresults}
As $l\rightarrow\infty$, the distance of the OFO solution from its optimal trajectory is given by:

\footnotesize
\begin{align}\label{convergence_asymptotic}
\lim_{l\rightarrow\infty}\psi(lm) \leq & \frac{1}{1-\epsilon_\mathrm{max}}\Big( |\Delta \bar{Q}_{g,\mathrm{max}}^\star| + \frac{2C_1C_3\alpha_qP_{g,\mathrm{max}}e^{-C_2}}{1-C_1e^{-C_2m}}\Big),
\end{align}
\normalsize where, $|\Delta \bar{Q}_{g,\mathrm{max}}^\star| = \max_{k} |\Delta \bar{Q}_{g}^\star(k)|$, $\epsilon_\mathrm{max}= \max_{k} |\epsilon(k)|$ with $\epsilon_\mathrm{max}<1$, $\alpha_q$ de-rating factor (see \eqref{maxpower}) and $P_{g,\mathrm{max}}$ maximum active power.
\end{corollary}
\begin{arxiv}
    \begin{proof}
        Refer to Appendix \ref{maintheoremproof}
    \end{proof}
\end{arxiv}
\begin{proof}
        \textcolor{black}{Refer to the proof in the extended version of our paper in} \cite{chandrasekaran2025reactive}.
\end{proof}
From the above Corollary, it can be observed that the asymptotic convergence of the OFO solution to the optimal trajectory of \eqref{optimizationproblem} is limited by the absolute maximum change in optimal reactive power between two OFO sampling times and the absolute maximum active power. Since $m=T_s/T_c$, we observe that convergence depends on the OFO sampling period $T_s$. A large $T_s$ promotes more time-scale separation between the inner-loop and the OFO controller, leading to the OFO trajectory being in close proximity to the optimal solution. However, a trade-off must be established, as an exceedingly large $T_s$ renders Assumption \ref{disturbance_assumption} impractical. 

\section{Experimental results}
\label{expRes}



\begin{figure}[!ht]
\centering{
\includegraphics[scale=0.3]{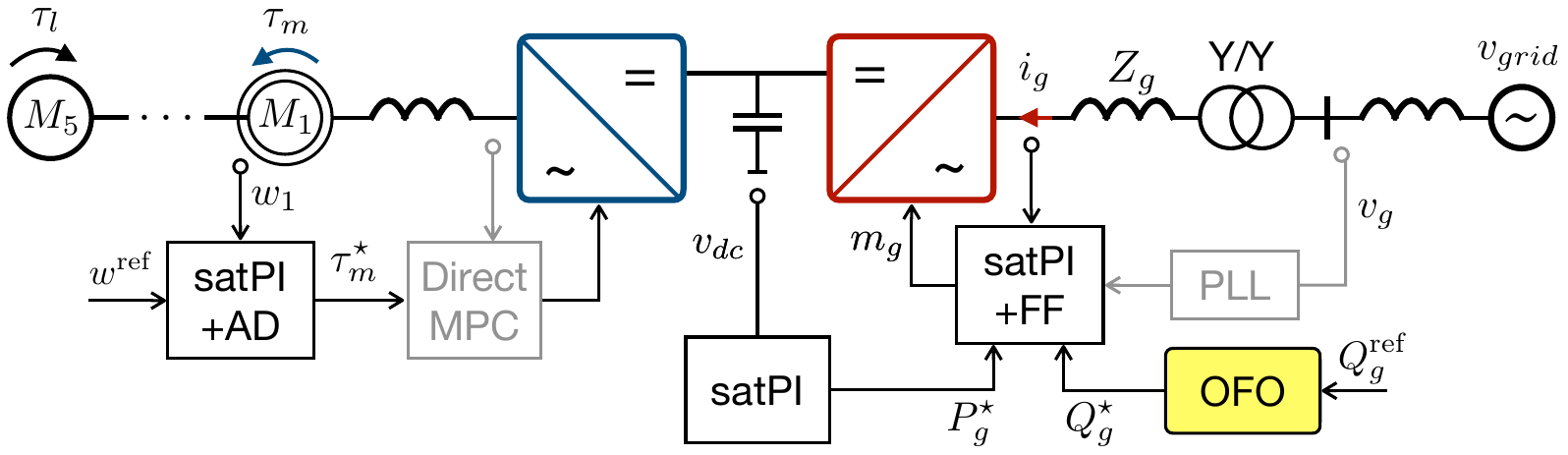}
\caption{Block diagram of the simulated setup. From left to right: speed control, torque control, voltage control and current control loops typically implemented in a drive system, together with a Phase-locked-loop (PLL) that extracts the grid voltage and the angle for the $dq$-transformation (as seen on the converter side of the transformer). The anti-windup tracking controllers are denoted by satPI, while further feed-forward (FF) and active-damping (AD) terms are appropriately added to improve the fidelity of the simulation.} 	\label{Fig: Block}}
\end{figure}

Through a proprietary software-in-the-loop setup, we simulate a pumped-hydro drive system with a 7MVA grid-side transformer 
and a 5kV DC-bus with mid-point. The converter is implemented using an average-model of the three-level active neutral-point-clamped (ANPC) topology. On the load-side, we implement a switching-model of the three-level ANPC converter connected to an Induction Machine (IM) model whose rotor is part of a 6MW driveline shaft modeled as a 5-mass torsional system. 
The air-gap torque $\tau_m$ acts on one end of the shaft and the load $\tau_l$ on the other. 
The motor torque is produced through (an industry-standard) direct Model Predictive Control (MPC) which takes $\tau_m$ as reference. As illustrated in Fig. \ref{Fig: Block}, the control structure described in Section \ref{control_section} is in place. 

All controllers, except for the OFO, were implemented with a sample time of $T_c = 250 us$ while the plant dynamics was discretized with a sampling time of $25us$. The modulation limit is set to $\tfrac{0.93}{\sqrt{2}}$. The two reactive power controllers 
were tuned as follows
\begin{itemize}
    \item Activation-function: $\omega_q = 2\pi5$, $\kappa_1 = 0.1$, $\kappa_2 = 250$, with soft-max $\Gamma_1$ at $i_{g,\mathrm{max}}$ and soft-max $\Gamma_2$ at $\tfrac{0.97}{\sqrt{2}}$;
    \item OFO: $k_\mu = 80$, $k_\gamma = 4$, $T_s = 10^{-3}s$, where $\mu = {k_\mu T_s}{\|v_g\|^2} \,\,,\, \gamma =  \tfrac{k_\gamma T_s}{\|v_g\|^2} \,$.
\end{itemize}

\begin{arxiv}
    \begin{figure}[!ht]
\centering{
\includegraphics[scale=0.4]{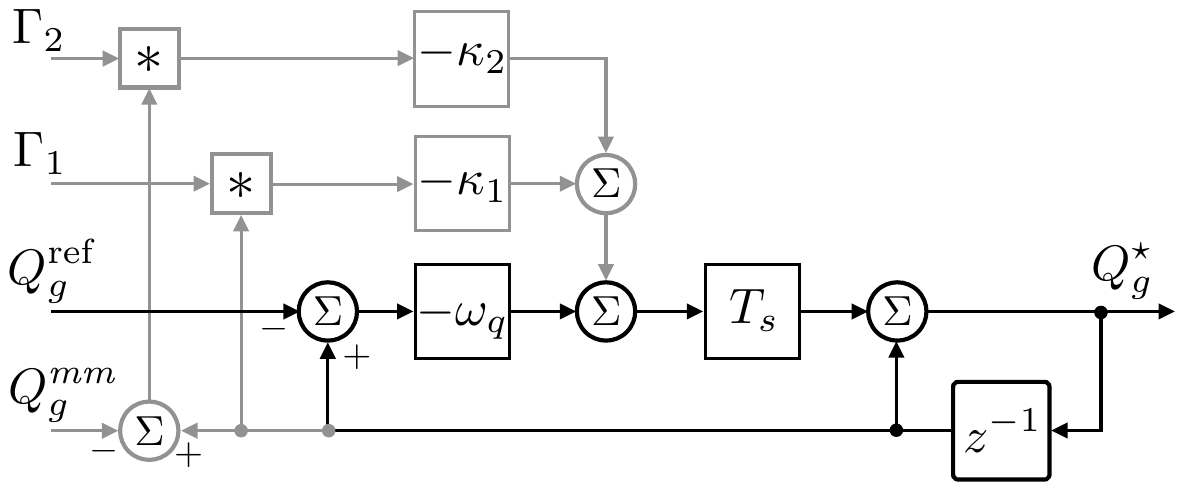}
\caption{Block diagram of the Forward Euler implementation of the activation function controller \eqref{AF_eqn}.}	\label{Fig: afblock}}
\end{figure}

\begin{figure}[!ht]
\centering{
\includegraphics[scale=0.4]{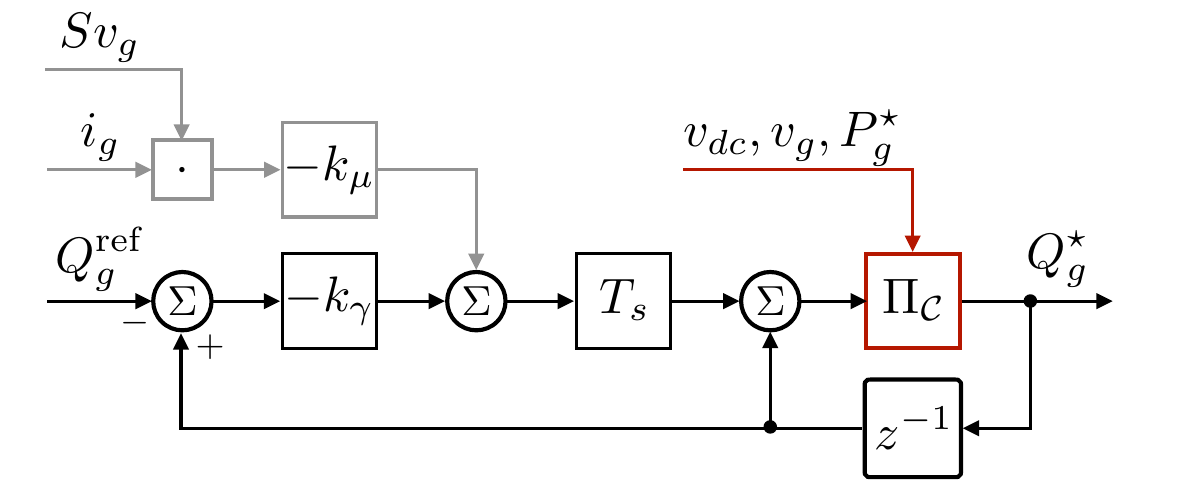}
\caption{Block diagram of the OFO algorithm implementation, where $S = [\begin{smallmatrix}0 & 1 \\ 1 & 0\end{smallmatrix}]$ is part of the sensitivity function in $\nabla\Phi$.}	\label{Fig: ofoblock}}
\end{figure}
\end{arxiv}


We perform three tests in which we compare the response of three scenarios: using (a) OFO, (b) Activation-function-based approach and (c) unmitigated, i.e. ${Q}_g^\star = {Q}_g^\mathrm{ref}$.

\subsection{Voltage dip}

\begin{figure}[!ht]
\centering{
\includegraphics[scale=0.4]{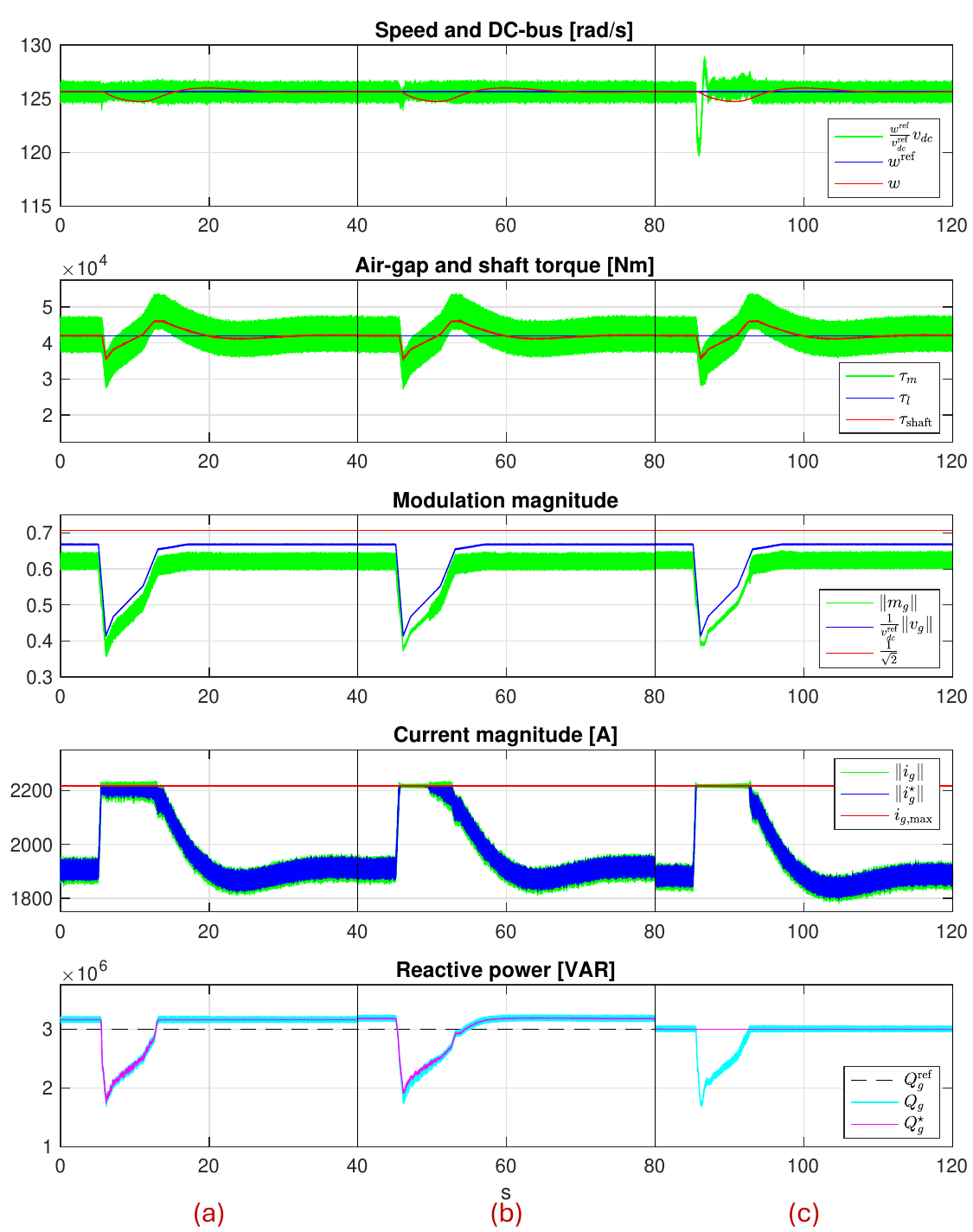}
\caption{Plots of the three responses: (a) OFO approach, (b) Activation-function approach, and (c) ${Q}_g^\star = {Q}_g^\mathrm{ref}$. In the first plot, we denote the rigid-body shaft speed $w$ in red, its reference in blue, and the DC-link voltage (scaled to units of speed) in green. In the second plot, we denote the air-gap torque $\tau_m$ in green, the load torque $\tau_l$ in blue, and a mid-shaft torque $\tau_\mathrm{shaft}$ in red. In the third plot, we denote the modulation amplitude $\|m_g\|$ in green compared to the magnitude of the grid voltage scaled by the DC-bus in blue, and the limit of the modulation amplitude in red. In the fourth plot, we denote the current magnitude in green, the reference magnitude in blue, and its limit in red. In the last plot, we denote the reactive power in cyan, its external reference in black, and the dynamic set-point in purple.}	\label{Fig: Dip}}
\end{figure}

We set the reference speed to $w_\mathrm{max}$ and the load to $0.9\tau_\mathrm{max}$, amounting to about 5.3 MW of load power, and we set the reactive power demand to 3 MVAR. We subject the drive system to a $40\%$ gradual dip in the amplitude of the grid voltage. 
%
%
Fig. \ref{Fig: Dip} shows the reactive power drop during transient, either driven by the set-point adaptation in cases (a) and (b), or induced by the balanced saturation of current reference magnitude in case (c). This test shows how the two proposed techniques produce an effect on reactive power and current magnitude similar to that of the circular current limiter but with direct intervention on the $Q_g^\star$ set-point rather than as a consequence of the saturation of $i_g^\star$.

\subsection{Reference step}

\begin{figure}[!ht]
\centering{
\includegraphics[scale=0.4]{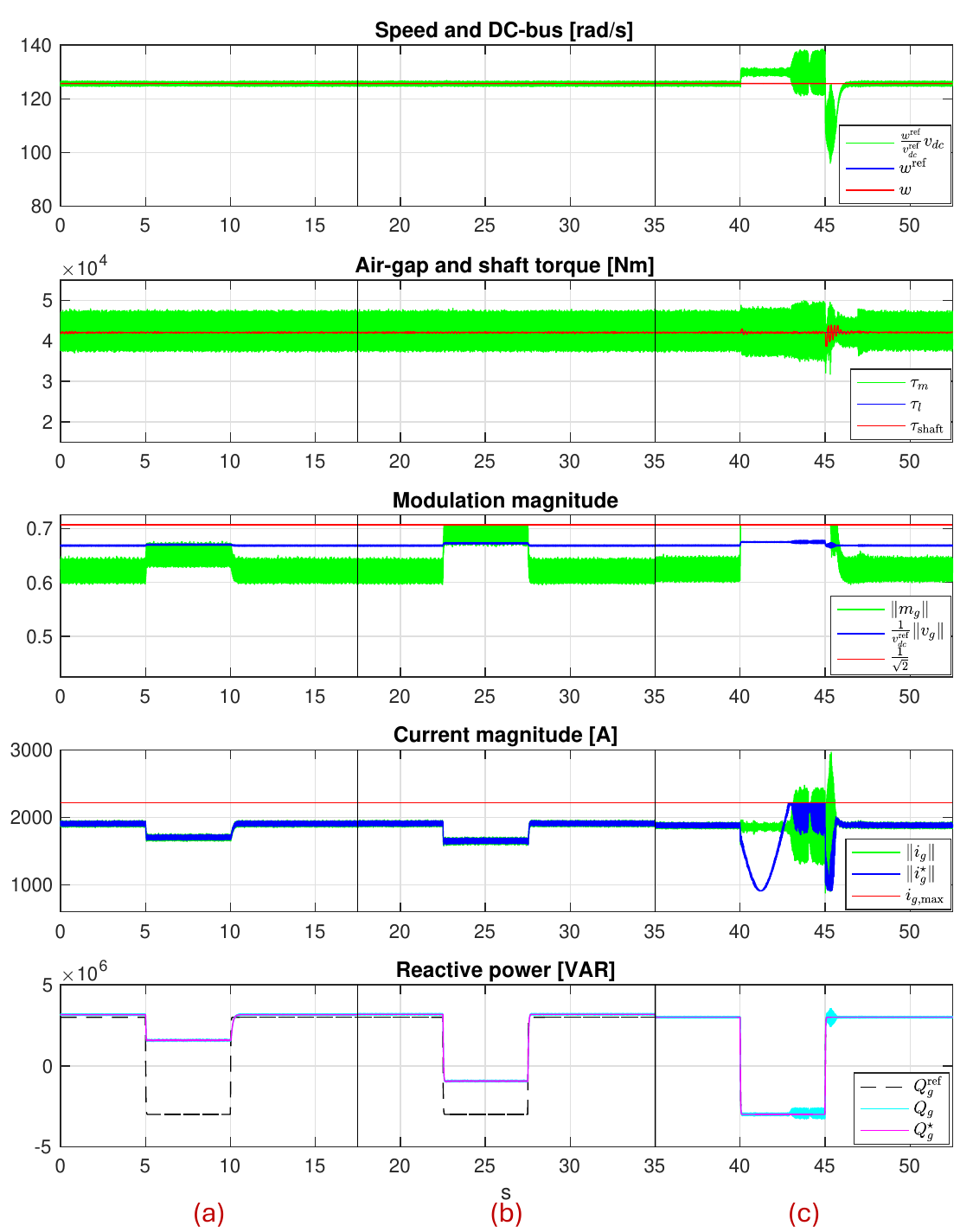}
\caption{Plots of the three responses to a reactive power demand outside of the constraint set: (a) OFO approach, (b) Activation-function approach, and (c) ${Q}_g^\star = {Q}_g^\mathrm{ref}$. Traces are as described in Fig. \ref{Fig: Dip}.}	\label{Fig: Step}}
\end{figure}

We apply a 5 second set-point change from 3 MVAR to (an infeasible) -3 MVAR, and back to 3 MVAR at the same load conditions as before. In Fig. \ref{Fig: Step}, we see how the two proposed methods (a) and (b) arrive at different trade-off between tracking the external reactive-power set-point and satisfying the modulation magnitude constraint. In the unmitigated case (c), we see how the current controller is no longer able to track the reference and goes into over-current (even though its reference is limited) as a consequence of modulation saturation. In addition, the DC-bus voltage loses regulation, reaches tripping levels, and disrupts motor-side operation. 


\subsection{Over-voltage}

We defer the presentation of our third test to the extended version of our paper in \cite{chandrasekaran2025reactive}.

\begin{arxiv}
    \begin{figure}[!ht]
\centering{
\includegraphics[scale=0.4]{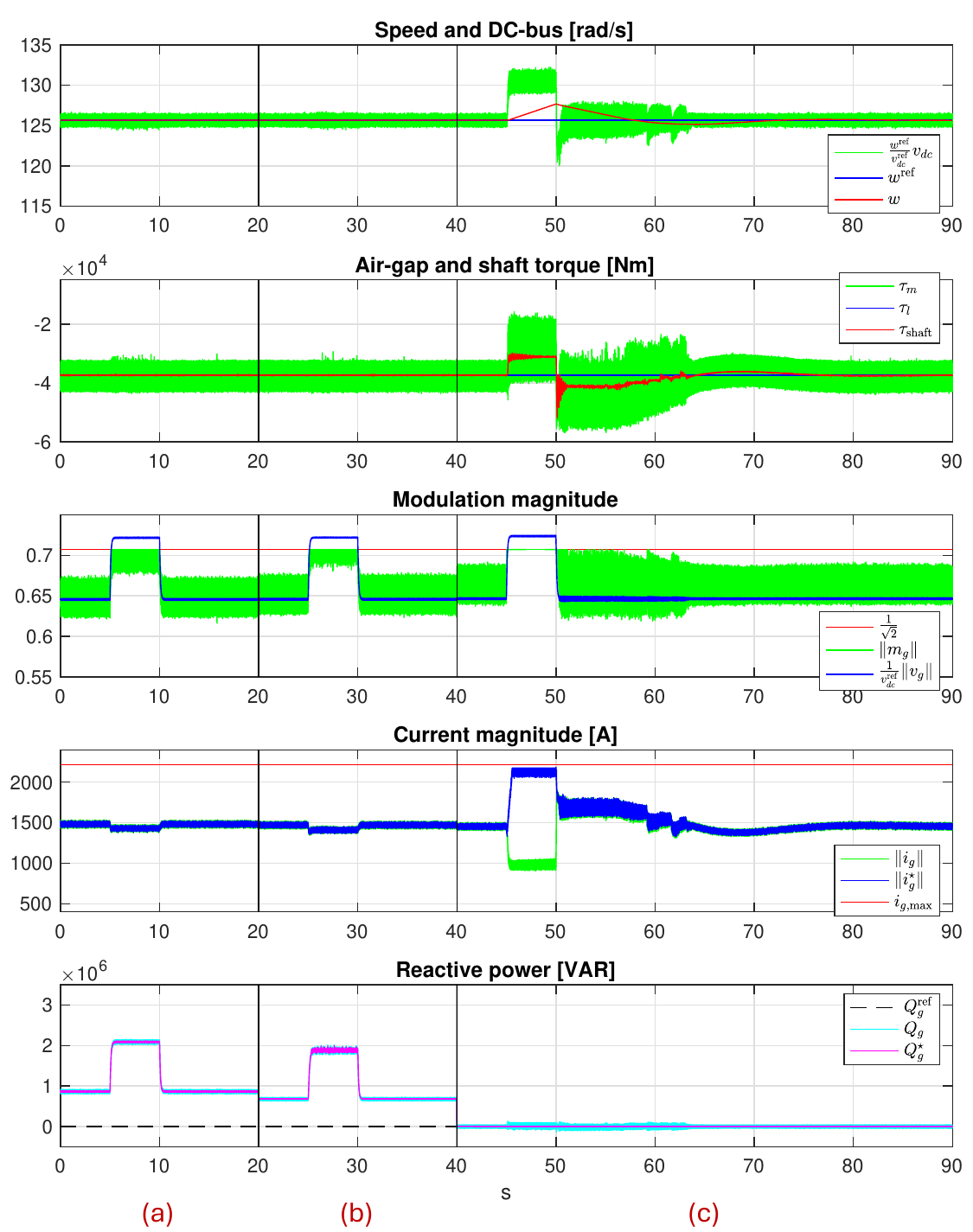}
\caption{Plots of the three responses to a severe grid overvoltage: (a) OFO approach, (b) Activation-function approach, and (c) ${Q}_g^\star = {Q}_g^\mathrm{ref}$. Traces are as described in Fig. \ref{Fig: Dip}.}	\label{Fig: OV}}
\end{figure}

In our third test, we leave the speed reference at $w_\mathrm{max}$ and set the load torque to $-0.8\tau_\mathrm{max}$. In this generator scenario, the reactive power reference is set to {\it zero}, and the grid undergoes a $12\%$ overvoltage for 5 seconds. We see in Fig. \ref{Fig: OV} that the proposed methods (a) and (b) produce slightly higher reactive power at steady-state, but ride trough the overvoltage. In the unmitigated case (c), the modulation saturates and produces an uncontrolled condition where a significant energy imbalance between motor-side and grid-side occurs. In this scenario, the (circular) current limiter is not in effect; in fact, not enough current is produced, and the motor-side energy is trapped in the driveshaft mass, causing it to overspeed. In addition, the DC-bus is overcharged and reaches tripping levels.
\end{arxiv}





\section{Conclusion}
\label{Concl}

This paper presents a novel approach to constraint-aware reactive power control in AC drives by leveraging the trade-off between input and output constraints. We draw the contrast between an activation-function method and an online feedback optimization approach.  
The proposed techniques are meant to complement the saturation mechanisms which are already in place, in order to improve the performance and extend the availability of the drive system by compromising reactive power tracking error. 
Our detailed simulation highlights the similarity of the two methods in mitigating system-wide fault conditions, showcasing their advantage over a conventional current limiter. 
%
%
%
That is, by maintaining the reactive power within the limits induced by the grid and load conditions, our approach avoids the saturation of the modulation vector, which in turn preserves the current tracking and the DC-bus regulation capability of the grid-side converter to the extent possible. As a result, the operation of the drive system is enhanced with minimum conservatism.



%

\begin{arxiv}
    \section
{Appendix}

\subsection{Proof of Theorem \ref{maintheorem}}\label{maintheoremproof}
We consider $\psi(k)=|Q_g^{\star}(k)-\bar{Q}_g^{\star}(k)|$, the distance between the OFO solution and the optimal trajectory as a performance criterion to show convergence guarantees as the problem \eqref{optimizationproblem} is strongly convex (see Remark \ref{convexityremark}).

Note that in the time instants where the OFO is not triggered, we have $Q_g^{\star}(k+1) = Q_g^{\star}(k)$. Further, since the disturbance is piece-wise continuous with respect to the OFO sampling rate (see Assumption \ref{disturbance_assumption}), the optimal trajectory $\bar{Q}_g^{\star}(k)$ remains constant between two OFO sampling times. Thus, the following relation can be obtained when the OFO is not triggered (i.e. $\zeta(k)=0$):
\begin{equation}\label{notriggereqn}
    \psi(k+1) \leq \psi(k).
\end{equation}
In the case where the OFO is triggered at time $k$ (i.e. $\zeta(k)=1$), we have the following inequality:
\begin{equation}\label{triggereqn}
    \psi(k+1) \leq |Q_g^{\star}(k+1)-\bar{Q}_g^{\star}(k)| + |\Delta \bar{Q}_g^{\star}(k)|,
\end{equation}
where, $\Delta \bar{Q}_g^{\star}(k) = \bar{Q}_g^{\star}(k)-\bar{Q}_g^{\star}(k+1)$ is the change in the optimal reactive power between two OFO trigger time instances. Note that the optimal reactive power satisfies the stationarity condition $\bar{Q}_g^{\star}(k) = \Pi_{\mathcal{C}(k)}[\bar{Q}_g^{\star}(k) - \mu(k)\Phi(\bar{Q}_g^{\star}(k),\bar{i}_g^{\star}(k))]$, where $\bar{i}_g^{\star}(k)= h(\bar{Q}_g^{\star}(k),d(k))$ and $\Phi(Q_g,i_g)$ is the composite gradient, as defined in \eqref{PGD}. We now use the projected gradient descent update on $Q_g^{\star}(k+1)$ and stationarity condition on $\bar{Q}_g^{\star}(k)$ in \eqref{notriggereqn}. We also use the fact that projection over a convex set is contractive (i.e. $\lVert\Pi_\mathcal{C}[x]-\Pi_\mathcal{C}[y]\rVert \leq \lVert x-y\rVert, \forall x,y\in\mathbb{R}^n$) to obtain

\footnotesize
\begin{align}
    \psi(k+1) \leq & \Big|(1-\gamma\mu(k))(Q_g^{\star}(k)-\bar{Q}_g^{\star}(k))- \mu(k)(H(k)^\top i_g(k+1) + \nonumber \\
    & \bar{H}(k)^\top\bar{i}_g^{\star}(k))\Big| + |\Delta \bar{Q}_g^{\star}(k)|\label{firsteqn} ,
\end{align}
\normalsize where, $H(k)=\nabla_{Q_g^{\star}}h(Q_g^{\star}(k),d(k))$ and $\bar{H}(k)= \nabla_{Q_g^{\star}}h(\bar{Q}_g^{\star}(k),d(k))$, the sensitivities of the steady-state map $h(Q_g^{\star},d(k))$ to $Q_g^{\star}$ at $Q_g^{\star}(k)$ and $\bar{Q}_g^{\star}$, respectively. Since the map $h(Q_g^{\star},d(k))$ is affine with respect to $Q_g^{\star}$ \eqref{ssmap}, $H(k) = \bar{H}(k), \ \forall k\geq 0$. Thus, \eqref{firsteqn} can be written as

\scriptsize
\begin{align}
    \psi(k+1) {\leq} & \underset{\mathcal{T}_1}{\underbrace{\Big|(1-\gamma\mu(k))(Q_g^{\star}(k)-\bar{Q}_g^{\star}(k))- \mu(k)H(k)^\top(i_g^{\star}(k)-\bar{i}_g^{\star}(k))\Big|}} + \nonumber\\
    & \underset{\mathcal{T}_2}{\underbrace{\Big|\mu(k)H(k)^\top(i_g(k+1)-i_g^{\star}(k))\Big|}} + |\Delta \bar{Q}_g^{\star}(k)|, \label{secondeqn}
\end{align}
\normalsize where we add and subtract the term $\mu(k) H(k)^\top i_g^{\star}(k)$ with $i_g^{\star}(k)= h(Q_g^{\star}(k),d(k))$ being the steady-state current at time $k$. Before we develop inequalities for terms $\mathcal{T}_1,\mathcal{T}_2$, we make the following observation:
\begin{remark}\label{ssmap_lipschitz}
    The steady-state map $h(Q_g^{\star},d(k))$ is Lipschitz continuous with respect to $Q_g^{\star}$ with Lipschitz constant $\lVert v_g(k) \rVert^{-1}$, i.e. $\lVert H(k)\rVert \leq \lVert v_g(k) \rVert^{-1} \ \forall k\geq0$
\end{remark}
We now focus on the term $\mathcal{T}_1$:

\scriptsize
\begin{align}
    \mathcal{T}_1^2 {=} & (1-\gamma\mu(k))^2(Q_g^{\star}(k)-\bar{Q}_g^{\star}(k))^2 + \mu(k)^2(H(k)^\top(i_g^{\star}(k)-\bar{i}_g^{\star}(k)))^2 \nonumber\\
    & -2\mu(k)(1-\gamma\mu(k))H(k)^\top(i_g^{\star}(k)-\bar{i}_g^{\star}(k))(Q_g^{\star}(k)-\bar{Q}_g^{\star}(k)) \nonumber\\
    \overset{(a)}{\leq} & (1-\gamma\mu(k))^2(Q_g^{\star}(k)-\bar{Q}_g^{\star}(k))^2 + \mu(k)^2\lVert v_g(k)\rVert^{-4}(Q_g^{\star}(k)-\bar{Q}_g^{\star}(k))^2 \nonumber\\
    & -2\mu(k)(1-\gamma\mu(k))H(k)^\top(i_g^{\star}(k)-\bar{i}_g^{\star}(k))(Q_g^{\star}(k)-\bar{Q}_g^{\star}(k)) \nonumber\\
    \overset{(b)}{\leq} & \Big((1-\gamma\mu(k))^2 + \mu(k)^2\lVert v_g(k)\rVert^{-4}\Big)(Q_g^{\star}(k)-\bar{Q}_g^{\star}(k))^2 \nonumber\\
    & -2\mu(k)(1-\gamma\mu(k))\lVert v_g(k) \rVert^{-2}(Q_g^{\star}(k)-\bar{Q}_g^{\star}(k))^2 \nonumber\\
    \leq & \Big(1-\gamma\mu(k) - \mu(k)\lVert v_g(k) \rVert^{-2}\Big)^2\psi(k)^2. \label{termT1} 
\end{align}
\normalsize In $(a)$, we make use of norm sub-multiplicativity, i.e. $(H(k)^\top(i_g^{\star}(k)-\bar{i}_g^{\star}(k)))^2 \leq \lVert H(k) \rVert^2\lVert i_g^{\star}(k)-\bar{i}_g^{\star}(k)\rVert^2$. We then make use of the Lipschitz property on $h(Q_g^{\star},d(k))$ to state that $\lVert H(k) \rVert^2 \leq \lVert v_g(k)\rVert^{-2}$ and $\lVert i_g^{\star}(k)-\bar{i}_g^{\star}(k) \rVert^2 \leq \lVert H(k)\rVert^{-2}(Q_g^{\star}(k)-\bar{Q}_g^{\star}(k))^2$. In $(b)$, we used the explicit analytical form for $H(k), i_g^{\star}(k),\bar{i}_g^{\star}(k)$. Thus, we have the term $\mathcal{T}_1 \leq |\epsilon(k)|\psi(k)$, where $\epsilon(k)= 1-\mu(k)(\gamma + \lVert v_g(k)\rVert^{-2})$. By choosing $\mu(k)\in\Big(0,\frac{2\lVert v_g(k)\rVert^2}{1 + \gamma\lVert v_g(k)\rVert^2}\Big)$, we ensure $|\epsilon(k)|<1, \forall \ k\geq 0$.

We now focus on the term $\mathcal{T}_2$.
\begin{align}
    \mathcal{T}_2 &\overset{(a)}{\leq} \mu(k)\lVert v_g(k)\rVert^{-1}\lVert i_g(k+1)-i_g^{\star}(k)\rVert \nonumber\\
    & \overset{(b)}{\leq} C_1 C_3\lVert v_g(k)\rVert\lVert i_g(k) - i_g^{\star}(k)\rVert e^{-C_2} \label{termT2}
\end{align}
In $(a)$, we make use of Lipschitz property on $H(k)$ to state that $H(k)\leq \lVert v_g(k)\rVert^{-1}$. In $(b)$, we make use of Assumption \ref{PI_assumption} to state that $\lVert i_g(k+1)-i_g^{\star}(k)\rVert\leq C_1\lVert i_g(k)-i_g^{\star}(k)\rVert e^{-C_2}$. Further, we choose $\mu = C_3\lVert v_g(k)\rVert^2$ and weight $\gamma(k) = C_4/\lVert v_g(k)\rVert^2$ with $C_3 \in(0,2/(1+\gamma v_{g,max}^2))$ for feasible guarantees and a more simplified convergence criterion with respect to the learning rate.

Using \eqref{termT1},\eqref{termT2} in \eqref{secondeqn}, we obtain
\begin{align*}
    \psi(k+1) \leq & |\epsilon(k)| \psi(k) + |\Delta \bar{Q}_g^{\star}(k)| + \\
    & C_1 C_3\lVert v_g(k)\rVert \lVert i_g(k) - i_g^{\star}(k)\rVert e^{-C_2}
\end{align*}
\normalsize Replacing $k$ with $lm$ in the above and using \eqref{notriggereqn} to state that $\psi((l+1)m)\leq \psi(lm+1)$, we obtain \eqref{convergence}. To prove asymptotic convergence, we trace \eqref{convergence} recursively till its initial conditions to obtain:
\begin{align}
    \psi(lm) \leq & \epsilon_{max}^l \psi(0)+ {|\Delta \bar{Q}_{g,max}^{\star}|}\sum_{t=0}^{l-1}\epsilon_{max}^{l-t-1} + \nonumber\\
        & {C_1C_3v_{g,max}e^{-C_2}\sum_{t=0}^{l-1}\epsilon_{max}^{l-t-1}\underset{\mathcal{T}_3}{\underbrace{\lVert i_g(tm) - i_g^{\star}(tm) \rVert}}} \label{recursivetracing}, 
\end{align}
\normalsize where, $\epsilon_{max} = \max_k|\epsilon(k)| < 1$ represents the contractivity rate and $|\Delta \bar{Q}_{g,max}^{\star}| = \max_k|\Delta \bar{Q}_{g}^{\star}(k)|$ is the maximum change in the optimal solution to \eqref{optimizationproblem} between two OFO sampling times. We now focus on the term $\mathcal{T}_3$ which is concerned with the transients in the inner loop:

\footnotesize
\begin{align}
    \mathcal{T}_3 \overset{(a)}{\leq} & \lVert i_g(tm) - i_g^{\star}((t-1)m)\rVert + \lVert i_g^{\star}((t-1)m)-i_g^{\star}(tm)\rVert \nonumber\\
    \overset{(b)}{\leq} & C_1e^{-C_2m}\lVert i_g((t-1)m) - i_g^{\star}((t-1)m)\rVert + 2i_{g,max} \nonumber\\
    \overset{(c)}{\leq} & \Big(C_1e^{-C_2m}\Big)^t\lVert i_g(0) - i_g^{\star}(0)\rVert + 2i_{g,max}\Big(\frac{1-(C_1e^{-C_2})^t}{1-C_1e^{-C_2}}\Big). \label{transients}
\end{align}
\normalsize In $(a)$, we add and subtract $i_g^{\star}((t-1)m)$ and use triangle inequality on the norm. In $(b)$, we use the exponential stability property in Assumption \ref{PI_assumption}, tracing till the previous OFO sampling time. Further, the current constraint in \eqref{constraintspace} ensures that $\lVert i_g^{\star}(k)-i_g^{\star}(m)\rVert \leq i_{g,max}, \forall k,m$. In $(c)$, we trace the recursive relation in $(b)$ to its initial conditions. In \eqref{transients}, we note that sufficient time-scale separation between the OFO and the inner control loop is required for convergence, i.e. $T_s$ must be sufficiently high so that $C_1e^{-C_2m}=C_1e^{-C_2T_s/T_c}<1$. By inserting \eqref{transients} in \eqref{recursivetracing} and using properties of geometric series at the limit of $l\rightarrow\infty$, we obtain the final asymptotic result in \eqref{convergence_asymptotic}.
\end{arxiv}

\bibliographystyle{IEEEtran}
\bibliography{CDC2025}

\end{document}